\documentclass[aps,showpacs,prl,toolkits,nofootinbib,preprint]{revtex4}
\usepackage{mathrsfs}
\usepackage{amsmath}
\usepackage{graphicx}
\usepackage{color}
\usepackage{epsfig}
\usepackage{subfigure}

\begin{document}

\title{  Tests of CPT   }

\author{Shmuel Nussinov}
\email{nussinov@post.tau.ac.il}
\affiliation{Tel Aviv University, Sackler School Faculty of Sciences, \\
Tel Aviv 69978, Israel\\  and \\
Schmid Science Center, Chapman University, \\
Orange, California 92866, USA}

\begin{abstract}

   The ongoing experimental efforts in the high energy and high precision communities keep
  providing evidence for CPT, a fundamental symmetry holding in any local Lorentz invariant
  theory.
  We suggest possible interconnections between different CPT violating parameters.
  Specifically, the very precise test of CPT in the $K^0-\bar K^0$ system suggests---though
  definitely does not imply---that CPT violations in other observable parameters (mass, width,
  charge, magnetic moments, etc.) are much smaller than the directly measured bounds.

\end{abstract}


\maketitle

\vspace{.25in}

     The discrete symmetries of $P$, the reflection of the three-space coordinates,
  $T$, time reversal, and charge conjugation, $C$, are not separately conserved.
Indeed both $ C$ and $ P$ are maximally violated by the charged current part of
the weak interaction Lagrangian. Also our present understanding of CP violations
implies that their apparent smallness reflects small CKM mixing
rather than any intrinsic, approximate, conservation.

However, CPT symmetry holds in all local, Lorentz invariant, field theories in
four dimensions.

The locality and ensuing analyticity of the $n$ point functions in momentum space
allow a ``Wick" rotation which is an analytic continuation to Euclidean space.
$PT$ then becomes the complete inversion, $ x_i \rightarrow - x_{i}, \; i=1,2,3,4$
  in Euclidean four-dimensional space. Unlike for odd  dimensions, this inversion
is a rotation and {\it not} a separate discrete transformation. Hence, $PT$ cannot
  be violated in a local Lorentz invariant theory of neutral scalar bosons.

   The fact that for complex spinors and other charged fields we need also $C$, charge
conjugation, and account for the anti-linear nature of time inversion to get the CPT
theorem is, however, highly  nontrivial.

CPT symmetry implies the equality of masses of particle and anti-particles, and for
unstable particles the equality of total widths $\Gamma$ and $\bar{\Gamma}$.
Also the equality of electromagnetic and other gauge couplings and ensuing magnetic
moments follow from CPT.

Some 40 experimental tests of CPT are listed in the PDG.\cite{PDG} Let $\delta_p(X)$
  be the precision with which the equality of property $p$ (mass, width, charge, magnetic
  moment, etc.) of a particle $X$ and the anti-particle $\bar{X}$ has been verified:
\begin{equation}
\delta_p(X) = |p(X)-p(\bar{X})|/p(X)
\label{deltapX}
\end{equation}

  The precisions vary over a very wide range and often are not better than  $\sim 10^{-5}$.
  In some cases where special efforts have been made,  e.g., the proton--anti-proton mass
  difference, $\delta_m(p) < 2 \cdot 10^{-9}$, was obtained.

  When $\bar{X}-X$ bound positronium-like states are available, some properties of the
  bound particle and anti-particle have been shown to be equal to within $10^{-12}$.

In one single case involving the $K^0-\bar{K}^0$ system stunning accuracies of:
\begin{equation}
   \delta_m(K^0)< 10^{-18}
\end{equation}

\begin{equation}
   \delta_{\Gamma}(K^0) < 10^{-17}
\end{equation}
has been achieved, reflecting the very well studied $K_L-K_S$ and $K^0-\bar{K}^0$
oscillations in this relatively long-lived system. In view of the importance of
the CPT theorem this is indeed most gratifying.  In passing we note that this system
allowed also precise tests of the equality of the gravitational couplings of $K^0$ and
  $ \bar{K}^0$, and a sensitive search for possible deviations from quantum mechanics.

  New studies of CPT conservation are underway. One example is measuring the equality
  of the top $(t)$ and $\bar{t}$ quark masses for which a precision of
$\delta_m(t) \sim 2.2$ was achieved.\cite{Abazov:2009xq}. This is quite
modest compared with the  more precise CPT tests. Yet it is worthwhile and may
be justified not only by the experimental challenge of finding the best bounds
  on any $\delta_p(X)$.
  Some models try to explain the large $m(t)$ by having the top quark experience
  effects of higher dimensional physics more than the other quarks so that some of
the assumptions underlying the proof of the CPT theorem may be slightly violated
there. Omitting such subtleties we would like to make the following observation:

   Barring unlikely fine-tuned cancelations, the very precise $\delta_m(K^0)$ and
  $\delta_{\Gamma} (K^0)$ suggest much stronger upper bounds on CPT  violations
  involving  quarks than what can be achieved via direct measurements.

Thus consider first  $\delta_m(t), \; \delta_m(c), \; \delta_m(u)$ and
$ \delta_m(W)$. Since weak interactions interconnect different quark (mass
eigenstates), any appreciable CPT violation in the $W$ or in the quark
masses ``trickles down" and affects $\delta_m(K), \;  \delta_{\Gamma}(K)$.

Thus consider the specific ``radiatively-induced"  difference of the $s $
and $\bar{s}$ quarks generated via charged current weak interactions
by the $t-W$ intermediate in the Feynman diagrams of Fig 1.
\begin{figure}[b]
\centering
\includegraphics[width=5in]{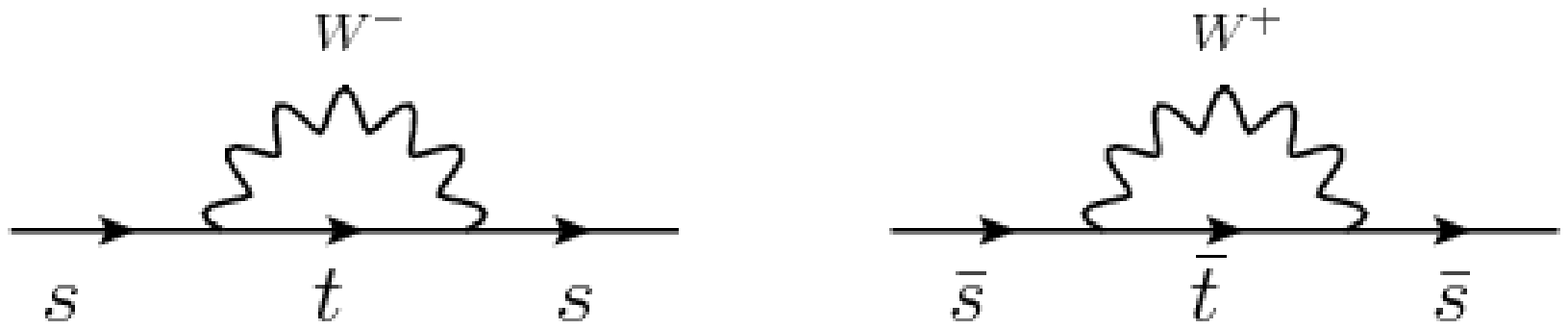}
\caption{  }
\label{CPTfig}
\end{figure}

\noindent Unlike each diagram  separately,  the difference is finite and
calculable.  It has the form:
\begin{equation}
\delta_m^1(s) = C(ts)\alpha_W/(2\pi) \; (V_{ts}^2 \;
                 m(t)-V_{\bar{t},\bar{s}}^2 \; m_{\bar{t}}) \nonumber\\
                 + C'(t,s)\alpha_W/(2\pi) \; (V_{ts}^2 \;
                 m(W)-V_{\bar{t},\bar{s}}^2 \; m(\bar{W})
\label{radinduced}
\end{equation}
with $C,C'$ pure numbers depending on mass ratios, the $V$'s elements of the
CKM  matrix and $\alpha_W$, the weak coupling at the relevant scale.

     The above difference can be rewritten as:
\begin{eqnarray}
\delta_m^1(s)& = &[\alpha_W]/[C[V^2 - \bar{V}^2](m_t+m_{\bar{t}}) \nonumber\\
              & + & [V^2 + \bar{V}^2](m_t-m_{\bar{t}})
                     +[\alpha_W]/[C'[V^2 -\bar{V}^2](m_W^- + m_W^+)]  \nonumber \\
              & + & [V^2 +\bar{V}^2](m_W^- - m_W^+)
\label{radinduceddiffCPT2.pdf}
\end{eqnarray}
This expression clearly separates between the contributions to $\delta_m(s)$
due to $t-\bar{t}$ (or $W^- - W^+)$ mass asymmetry and those generated by
different $V =V_{t,s}$ and $\bar{V} = V_{\bar{t},\bar{s}}$ CKM mixings. Similar
expressions can be written for the contribution of the $c-W$ and of $u-W$
intermediate states, generating altogether 12  terms.

    Barring unlikely cancelations between these 12 different terms contributing
  to the strange quark  and the anti-quark masses we can use the small
  $\bar{s}-s$ mass difference to bind any of the above terms:
\begin{eqnarray}
   m_s-m_{\bar{s}} & > & V_{t,s}^2 \; \alpha_W/(2\pi) (m_t-m_{\bar{t}}) \\
   m_s-m_{\bar{s}} & > & V_{c,s}^2 \; \alpha_W/(2\pi) (m_c-m_{\bar{c}}) \\
   m_s-m_{\bar{s}} & > & V_{u,s}^2 \; \alpha_W/(2\pi) (m_u-m_{\bar{u}})
\label{bindedterms}
\end{eqnarray}
and
\begin{equation}
m_s-m_{\bar{s}} > [V_{us}^2 + V_{cs}^2 + V_{ts}^2](m_{W^-} - m_{W^+}) =
m_{W^-} - m_{W^+}
\end{equation}

Such considerations apply also to the mass difference of the $\bar{d}$ and
$d$ quarks yielding the analogs of all the above relations with
$d \leftrightarrow s$ everywhere.  Thus we have
\begin{equation}
   m_d - m_{\bar{d}} > V_{t,d}^2 \; \alpha _W/(2\pi) (m_t - m_{\bar{t}}),
\;\;{\rm etc.}
\label{analogs}
\end{equation}
Similar bounds are obtained for mass weighted asymmetries of the CKM matrix
elements for quarks and anti-quarks.

      In addition to the above ``flavor off-diagonal contribution" to the
mass differences of the $d$ and $s$ quarks and corresponding anti-quarks,
we have the $Z$, photon and gluon exchange contributions.

      The next and last step in connecting with the experiment is to find
how a putative $\delta_m(s)$ and/or $\delta_m(d)$ reflects in the measured
  $\delta_m(K)< 10^{-18}$.

Strictly speaking, this involves nonperturbative QCD and may require lattice
calculations of mesonic ``$\sigma$ terms".  For a rough estimate we use the
expression for the masses of Nambu-Goldstone boson in terms of bare
quark masses:\cite{GMOR}
\begin{eqnarray}
   m^2(K^0) &  = & [m^0(s) +m^0(d)] <\bar{q} q>/(f^2(K^0))  \nonumber \\
   m^2(\bar{K}^0) & = &  [m^0(\bar{s}) +m^0(\bar{d})] <\bar{q}q>/(f^2(\bar{K}^0))
\label{gmorEq}
\end{eqnarray}

By subtracting these two equations we relate $\delta_m(K)$ to $\delta_m(s)$ and
  $\delta_m(d)$ (and to the difference of decay constants $f(K^0)-f(\bar{K}^0)$.
   Again assuming no cancelations and using
<$\bar{q} q> \sim $ (300 MeV)$^3$,  $f(K) \sim $  150 MeV, $\alpha_W \sim 1/(20)$
and the values of quark masses and mixing  parameters from the PDG we
finally obtain:
\begin{eqnarray}
   \delta_m(t) & = & [m_t - m_{\bar{t}}]/(m_t) \nonumber \\
               & = & \delta_m(K)[m(K)/(m(t))]2\pi/(\alpha_W)V_{ts}^{-2}
               \sim 10^{-18}V_{ts}^{-2} = 10^{-15} \nonumber \\
   \delta_m(W) & = & [m_{W^-} -m_{W^+}]/(m_W) \nonumber \\
   \delta_m(K) & = & [m(K)/(m(W))]2\pi/(\alpha_W)/([V_{ts}^2+V_{cs}^2+V_{us}^2])
                \sim 10^{-18}
\end{eqnarray}
(See footnote.)\footnote{The last inequality may be weaker by two orders of
magnitude due to the following:  To have non-vanishing radiative corrections to
the quark masses, we do need some bare quark masses breaking the chiral invariance.
Thus while the $W^+ - W^-$ mass difference is essential for the CPT asymmetry in the above terms,
the radiative corrections may involve an extra smaller, say, $c$ quark mass
of $\sim$ 1 GeV rather than 100 GeV.}
\begin{equation}
   delta_m(c)  \sim 10^{-16}/(V_{cs}^2) \sim 10 ^{-16}
\end{equation}

   In the above we implicitly assumed that QCD-gluon exchange interactions
are equally strong for the $\bar{s}-d$  and $\bar{d}-s$ systems.  Hence
using the same line of reasoning and excluding accidental cancelations between CPT
violations in SU(3)$_c$ couplings and in bare quark masses we conclude also that:
\begin{equation}
   g_{\bar{s},d} g_d  - g_{\bar{d},s}g_s < 10^{-18}
\end{equation}

All the above exceed by many orders of magnitude the direct bounds for
these asymmetries now and in the foreseeable future.

  Many other bounds, in particular, on differences between mixings of quarks
and  of anti-quarks in the CKM  and ``$\bar{CKM}$ matrix"  are suggested
  by the  almost equally stringent bound on the difference of widths:
\begin{equation}
   \delta_{\Gamma}(K^0) < 10^{-17}
\end{equation}
We will not discuss these relations here.

  It is useful to elaborate a bit more on the philosophy underlying this note.
  Altogether $\delta_m(K^0)$ can be expressed as a sum of $\sim$ 20 terms
depending on quark masses and gauge coupling, where each term violates CPT.
Since there is no separate control of each term, an almost complete
cancelations of relatively large CPT violations in each of these terms
in the overall sum cannot be excluded.

  We note, however, that the CPT theorem holds not only in
our specific ``standard model" with its $\sim$ 17 independent parameters,
but in any other local--standard model-like--field theory. Thus we can
vary the strength of the gauge couplings, the mass scales of QCD and of the
weak interactions, and the various diagonal and off-diagonal couplings of the
Higgs particle to quarks and leptons and CPT should hold equally well.

  Phrased differently, all observable CPT violations should trace back to CPT
  asymmetries in these fundamental, underlying parameters or any other set
of parameters in any yet more fundamental underlying theory.  Hopefully such
a more fundamental theory will have a smaller number of independent parameters.

  The theory must deviate in some way from purely local Lorentz invariant field theory
  so as to allow for violation of the CPT theorem. Most likely all CPT violations
will then trace back just one or very few novel features such as some form of non-
locality.  In such a scenario it seems extremely unlikely that the one or two
sources of CPT violation will be large so as to yield large CPT asymmetries in each
of the above $\sim$ 20 terms and yet conspire to have the incredible precise
cancelations in $\delta_m(K^0)$.

    In some ``landscape approaches"  it is believed that we live in one particular
  string theory vacuum with many of the specific SM  parameters fixed by anthropic
  considerations. In any case we cannot perform the above Gedunken experiment of
dialing  the various parameters and  verify that the measured $\delta_m(K)$
  is equally small in all cases.

    Having no reason to believe that a larger CPT violation in the $K-\bar{K}$ system, say,
  $ \delta_m(K) \sim 10^{-10}$, would prevent intelligent life, it seems unlikely
  that our vacuum conspired to minimize $\delta_m(K)$.  Assuming then that the various
  terms contributing to  $\delta_m(K)$ have random relative signs the probability that
  their sum will be so small unless these are separately small is truly tiny. Hence
  we adopted the ``No fine-tuned cancelation" hypothesis and looked for its consequences.

      Finally we would like to briefly comment on the equality of the electron and (minus)
  the positron charges. The direct bound quoted in the PDG is:
\begin{equation}
   \delta_q(e) = (q_{e^-} +q_{e^+})/(|q_e|) < 4 \cdot 10^{-8}
\end{equation}
A much better indirect bound can be obtained if we assume electric charge conservation
  as follows:

     Charge conservation in the annihilation: $e^+ e^- \rightarrow \gamma+\gamma$
  implies that  $\delta_q(e) = q(\gamma)/(|q_e|)$.  The charge of photons is, however,
strongly bound,  $q(\gamma) < 10^{-33}(|q_e|)$!  by using the coherence
in the Brown,
Hanbury,  Twiss effect measuring the relative phase between two paths of a photon.

For a charged particle this phase is the appreciable Aharonov-Bohm phase
$\sim$ to the total flux enclosed; hence, the very strong  bound.\cite{Altschul:2007xf}

\medskip

I would like to thank Tom Ferbel for telling me about the recent measurement of
top and anti-top masses which inspired this short note.

\end{document}